    \documentstyle[nato,epsfig,numreferences]{crckapb}

\begin{opening}
\title{HST - WFPC2 PHOTOMETRY OF THE GLOBULAR CLUSTER \protect\\
NGC 288: BINARY SYSTEMS, BLUE STRAGGLERS AND VERY BLUE STARS}

\author{M. Bellazzini}
\author{M. Messineo}
\institute{Osservatorio Astronomico di Bologna\\
           Via Ranzani 1, 40127 Bologna, ITALY}
           
\end{opening}

\runningtitle{WFPC2 OBSERVATIONS OF NGC 288}

\begin{document}

\begin{abstract}
We report on new WFPC2 observations of the globular cluster 
NGC 288, focusing our attention on peculiar stars. A very pronounced binary
sequence, paralleling the ordinary Main Sequence (MS) is clearly observed in 
the Color Magnitude Diagram (CMD) and a huge relative fraction of Blue
Straggler Stars is measured. The dataset offers the opportunity of studying the
evolution of a large population of binaries (and binary evolution by-products)
in a low density environment, where the evolution of such systems is not
dominated by collisions and encounters. 
Three (very) Extreme Horizontal Branch Stars have been found, 
all lying outside of the cluster core. 

\end{abstract}

\section{Introduction}

As a part of a long term programme aimed to obtain complete samples of all the
evolved stars in a few prototypical globular clusters\footnote{The programme is
named {\em Large Population Studies of Globular Clusters (LPSGC)} and is a
collaboration between scientists from the Bologna and Rome observatories, 
the Goddard Space Flight Center (Baltimore, MD) and the University of Virginia 
(Charlottsville), see Ferraro, this volume.}, we have obtained multi-wavelenght 
observations of a couple of WFPC2 fields towards two very important globulars:
NGC 2808 and NGC 288. 

Here we report some preliminary result about NGC 288, a cluster that holds a
key role in the literature about the {\em Second Parameter} phenomenon (see
\cite{ffpb} for a review, and references therein). 

\section{Observations and data reductions}

We observed a field centered on the core of the cluster (Central Field,
hereafter CF) with the filters
F255, F336, F555, F814 and a field a couple of arcmin apart (Outer Field,
hereafter OF) with the filters F555 and F814. Because of a problem in the 
pointing the two fields have a small overlapping area and the OF observations
saturates $\sim 0.5 ~mag$ above the Horizontal Branch (HB) level.

The data reduction has been performed with DOPHOT \cite{dop}, according to
the general prescriptions adopted by \cite{ols}. The absolute calibration in the
Johnson-Cousins system has been obtained adopting the Holtzman relations
\cite{holtz}.
The final magnitudes are the average of various (2-4) measures on repeated
exposures. 

\section{The CMD: a clear secondary sequence}

In fig. 1 the V {\em vs.} V-I CMD of the whole sample and the V {\em vs.} U-V
CMD of CF are shown. The most striking feature of these diagrams is the evident
secondary sequence paralleling the Main Sequence that is naturally associated
with a population of couples of MS stars measured as a single source. This
occurrence can be due to chance superposition of otherwise unrelated stars
or to the physical association of the couple (i.e. binary systems,
see \cite{hut} for a review).

{\small
\begin{figure}
\epsfig{file=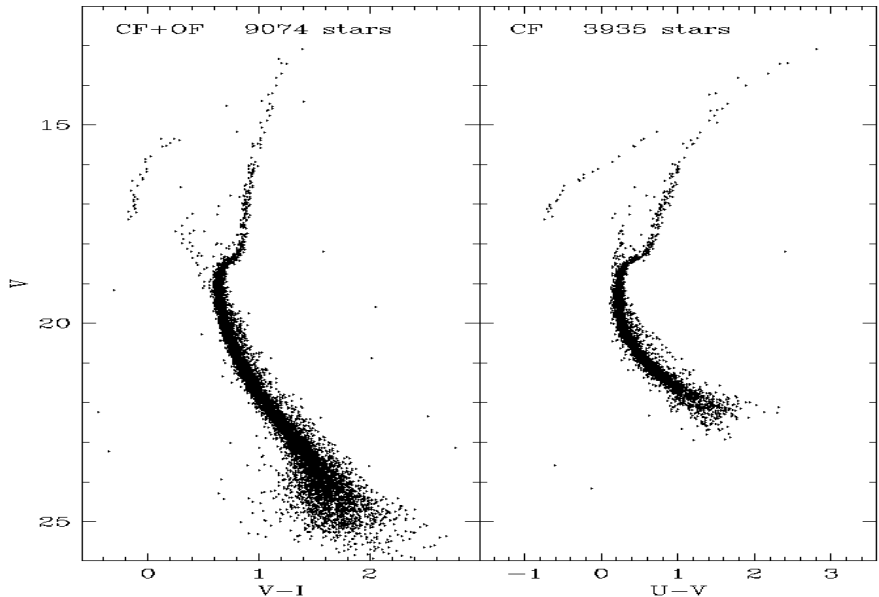, height=8truecm, width=12truecm}
\caption{(V-I,V) and (U-V,V) Color magnitude diagrams of NGC 288.}
\end{figure}
}
 
To evaluate the number of chance blending contaminating the observed secondary
sequence and to obtain robust estimates of the true binary
fraction, extensive artificial stars experiments are needed \cite{rub} and
are presently in progress. However we are dealing with a very low density and
relatively nearby cluster, so crowding has not to be a major concern in this
case. The probability that two stars images overlap to within 2 FWHM is
$P_{overlap}=\pi*FWHM^2*N_{s}/A$, where $N_s$ is the number of stars in the
observed field and $A$ is the area of the field. For our Central Field,
the most ``crowded'' one, $P_{overlap} < 3 \%$, thus the contamination by chance
blending is expected to be negligible. So, it can be safely expected that most
of the stars populating the secondary sequence are genuine binary systems and
that the binary fraction in this cluster is remarkably high.

\section{Blue Stragglers: a very high fraction}

It is now widely accepted that Blue Straggler Stars (BBS) are products of the
evolution of binary systems \cite{bai}. If the above discussed secondary
sequence is indeed due to a large binary fraction in NGC 288, a
significant population of BSS is also expected. This is actually the case. 
In fig. 2 the BSS are clearly identified in the F255 {\em vs.}
F255-U CMD. While the absolute number of BSS is not particularly remarkable
(as expected for such a low mass cluster), the specific frequency - defined by
\cite{ferbss} as the ratio of the number of BSS and the number of HB stars  -
is one of the highest ever found in a globular, $F_{HB}^{BSS}=1.35$.
In table 1 \footnote{First three columns from \cite{D93}, last column
from \cite{ferm80} and the present work.} the $F_{HB}^{BSS}$ ratio is reported 
together with other relevant
parameters for a handful of clusters observed with HST-WFPC2 and with the
adoption of the same selection criteria for the BSS sample (see \cite{ferm80}).
The sample is very small but homogeneous and fully self-consistent for what
concern the census of BSS.

{\small
\begin{table}
\begin{center}
\caption{Comparison between the $F_{HB}^{BSS}$ and other relevant parameters
for an homogeneous selected sample of Globular Clusters. The clusters are
ordered with growing central density.}
\begin{tabular}{lllll}
\hline
ID & [Fe/H] & log${\rho_0}_{L_{\odot V} pc^{-3}}$ & ${t_{rc}}_{[10^9yr]}$&
$F_{HB}^{BSS}$ \\
\hline
NGC 288       &-1.40&1.80&1.230&1.35\\
NGC 6205 - M13&-1.65&3.32&0.871&0.17\\
NGC 5272 - M3&-1.66&3.54&0.602&0.67\\
NGC 6341 - M92&-2.24&4.38&0.069&0.55\\
NGC 6093 - M80&-1.64&4.79&0.040&1 - 1.7\\
NGC 7099 - M30&-2.13&5.07&0.002&0.49\\
\hline
\end{tabular}
\end{center}
\end{table}
}

It is immediately evident that the only cluster with a comparable $F_{HB}^{BSS}$
ratio is M80 which is a very dense cluster and that harbours most of its BSS in
the very central regions. In particular the two entries for $F_{HB}^{BSS}$
refer to the global and the central value respectively. \cite{ferm80} argues
that the exceptional (and exceptionally concentrated) BSS population in M80 
can be understood if it is assumed that the cluster is in a very special phase 
of its dynamical evolution, i.e. it is struggling against the unavoidable 
core collapse, a favorable condition for the formation of {\em collisional}
BSS. However, while the collision time (see \cite{bintre}, eq. 8--125) 
for a MS star in the core of M80 is $t_{coll}(M80) \sim 5 ~Gyr$, 
$t_{coll}(NGC288) \sim 1000 ~Gyr$, so collisions between MS
stars has to be exceedingly rare. Thus, it is very likely that the large
majority of BSS in NGC 288 originated from a population of
{\em primordial binaries} through a mechanism as efficient as the one operating
in M80, but completely different.

{\small
\begin{figure}
\epsfig{file=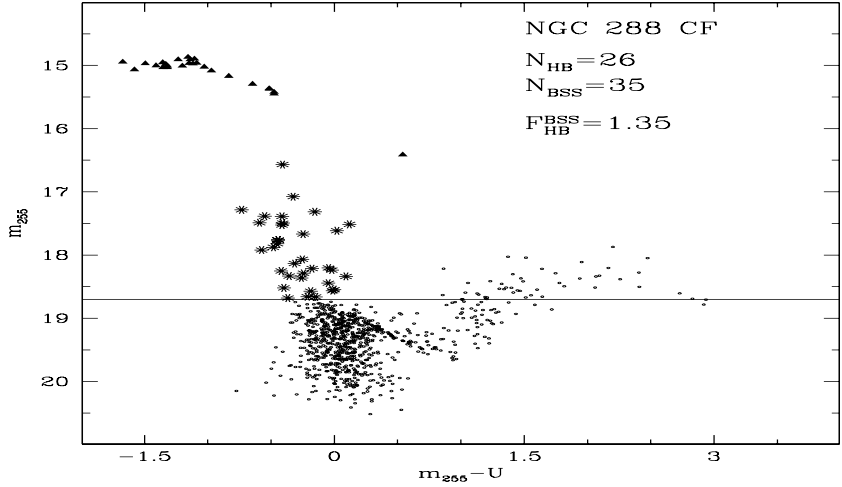, height=7truecm, width=12truecm}
\caption{$F_{HB}^{BSS}$ ratio from the selection operated in the F255W {\em vs.}
F255W-U plane, according to the method adopted by Ferraro et al. (1999). BSS
are indicated by asterisks and HB stars by  triangles. The faint limit
for BSS selection is indicated by the $m_{255}=18.8$ line.}
\end{figure}
}

\section{HB morphology: Extreme Horizontal Branch stars?}

In the right panel of fig. 3, a zoomed view of the HB stars in the (V, V-I)
CMD is reported, while the left panel report the position of the same stars
within the whole observed field. Three main facts can be remarked about the HB
morphology: 

{\small
\begin{figure}
\epsfig{file=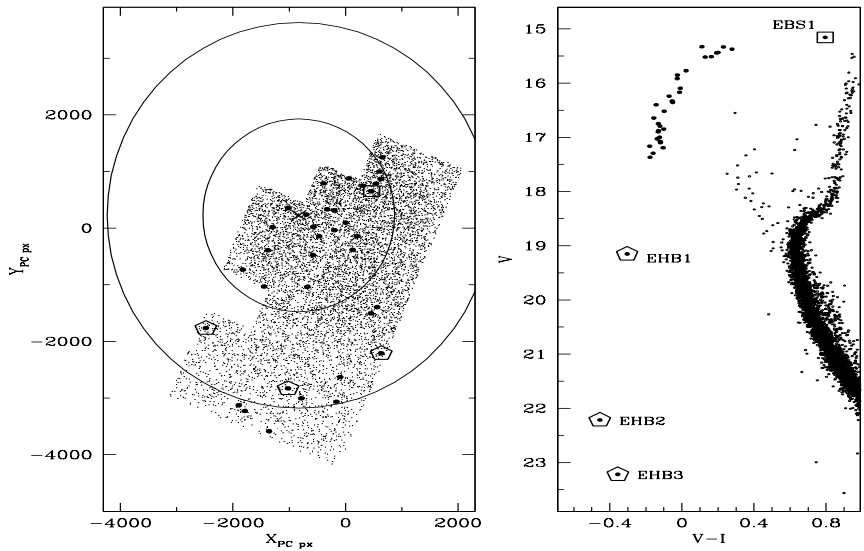, height=8truecm, width=12truecm}
\caption{{\bf Right panel:} HB morphology for the CF+OF sample. HB stars are
indicated by larger dots, the candidate Evolved BSS by the open square and the
candidates EHB stars by pentagons.{\bf Left panel:} Map of the whole sample.
North is up and East is Left, the symbols are the same as in the right panel.
The cross marks the center of the cluster. The circles with radius of 1 $r_c$
and 2 $r_c$ are also reported.}
\end{figure}
}

\begin{itemize}

\item The overall HB morphology is the well known - almost purely blue - one of
NGC 288. There is no possible candidate RR Lyrae star and a marginally
significant gap around $V-I=0.05$.

\item There is a single red HB star. Since the red clump is the place where the
Evolved BSS (EBSS, \cite{fusfer}) are expected to lie, during their core He 
burning phase, may be that this anomalous HB star is indeed an EBSS. Based on
crude evolutionary time consideration one would expect $N_{EBSS}\sim
0.6*N_{BSS}$. In the present case $N_{EBSS}\sim 2$ is expected, in good
agreement with the observed sole red HB star. Note that the EBSS candidate is
found near the center of the cluster (fig. 3, left panel).

\item There are three very blue stars on the ideal extension of the HB
ridge line, much fainter than the end of the main HB distribution. All of these
Extreme HB (EHB) candidates are genuine, isolated stars that have many 
consistent measurement in each filter from repeated exposures.
All of them lie well beyond the cluster core, far from the densest region of
the cluster. Unfortunately we have only V and I observations for these stars
since they all fall in the OF. While EHB1 is fully compatible with being a
bona-fide HB star, the nature of EHB2 and EHB3 is much more uncertain. It may
be that no reasonable HB model would be able to push stars at such high
temperatures. Tests in this sense are in progress and other hypothesis are
under consideration. Preliminary checks seems to exclude background quasars,
field stars, white dwarfs and ordinary cataclysmic variables as viable
possibilities.

\end{itemize} 

\section{Acknowledgments}

The authors are grateful to all the members of the LPSGC team: F. Fusi
Pecci, F.R. Ferraro, C. Cacciari, R. Buonanno, C.E. Corsi, G. Marconi, R.T.
Rood, B. Dorman and W. Landsman. 
This research is partially funded by a MURST Grant assigned to the project
``Stellar Evolution'' (national coordinator prof. V. Castellani).

\end{document}